\documentstyle[12pt]{ioplppt}

\begin{document}

\title{Orientational phase transitions in anisotropic rare-earth magnets at low
temperatures}
\author{V.Yu. Irkhin$^{*}$ and A.A. Katanin}
\address{620219, Institute of Metal Physics, Ekaterinburg, Russia}

\begin{abstract}
Orientational phase transitions are investigated within the Heisenberg model
with single-site anisotropy. The temperature dependence of the cone angle is
calculated within the spin-wave theory. The role of the quantum
renormalizations of anisotropy constants is discussed. A comparison with the
experimental data on the cone-plane orientational transition in holmium is
performed.
\end{abstract}

\maketitle

\section{Introduction}

The old problem of magnetic structure of rare-earth metals and their
compounds is still a subject of experimental and theoretical investigations.
These substances have complicated phase diagrams and demonstrate a number of
orientational phase transitions. In particular, such transitions take place
in the orthoferrites and practically important intermetallic compounds RCo$%
_5 $ (R=Pr,Nd,Tb,Dy,Ho), see, e.g., Ref. \cite{Levitin}. Qualitative
explanation of these transitions has been obtained many years ago within the
Heisenberg model with inclusion of magnetic anisotropy \cite{Coqblin}. In a
number of systems, lattice (magnetoelastic) effects are important. The
standard description is usually performed within mean-field approaches.
However, quantitative comparison with experimental data requires a more
detailed treatment.

Provided that the orientational transition temperature is low (in comparison
with the magnetic ordering point), spin-wave theory is applicable \cite
{Coqblin}. In the simplest case of the second-order anisotropy the
magnetization lies either along the easy axis or in the easy plane.
Inclusion of higher-order anisotropy constants can lead to cone phases where
magnetization makes the angle $\theta $ with the $z$-axis. The case $\theta
=\pi /2$ was considered in Refs. \cite{Tb,TbDy1,GdTbDy,TbDy2} where the
temperature renormalization of anisotropy constants and the spin-wave
spectrum in Tb and Dy within the standard spin-wave theory were calculated.

In the present paper we consider the cone phase with arbitrary $0\leq \theta
\leq \pi /2$. The situation here is analogous to that for the field-induced
orientational phase transitions, e.g., in the transverse-field Ising model
(see Ref.\cite{OurITF} and references therein). Unlike the latter model, one
can expect that at low enough temperatures and small value of anisotropy the
spin-wave theory is applicable for an arbitrary relation between anisotropy
parameters, not only close to the orientational phase transition. Even for
the second-order easy-plane anisotropy, the Holstein-Primakoff
representation for spin operators used in Refs. \cite{Tb,GdTbDy,TbDy2} leads
to so-called kinematical inconsistencies because of incorrect treating
on-site kinematical relations. To avoid this difficulty, we use the
technique of spin coherent states. Our approach is to some extent similar to
the operator approach used in Ref. \cite{TbDy1}, but gives a possibility to
treat more simply higher-order anisotropy constants, as well as to calculate
higher-order terms of $1/S$-expansion.

The anisotropic Heisenberg model used is formulated in Sect 2. In Sect 3 we
develop a special form of the $1/S$-expansion which gives a possibility to
take into account exactly on-site kinematical relations. In Sect 4 we treat
the cone-plane transition owing to the temperature dependence of the cone
angle $\theta $ and discuss experimental data on the rare earth metals.

\section{The model and mean-field approximation}

We start from the Hamiltonian of the Heisenberg model with inclusion of
single-site anisotropy
\begin{equation}
{\cal H}=-\frac J2\sum_{\langle ij\rangle }{\bf S}_i{\bf S}%
_j+B_2^0\sum_i(O_2^0)_i+B_4^0\sum_i(O_4^0)_i  \label{H}
\end{equation}
where $J>0$ is the exchange parameter,
\begin{eqnarray}
O_2^0 &=&3(S^z)^2-S(S+1)  \nonumber \\
O_4^0 &=&35(S^z)^4-30S(S+1)(S^z)^2+25(S^z)^2  \nonumber \\
&&\ +3S^2(S+1)^2-6S(S+1)  \label{Tens}
\end{eqnarray}
are the irreducible tensor operators of second and fourth orders, $B_l^m$
are the corresponding anisotropy constants.

Up to unimportant constant we can rewrite the Hamiltonian (\ref{H}) in the
form
\begin{equation}
{\cal H}=-\frac J2\sum_{\langle ij\rangle }{\bf S}_i{\bf S}%
_j+D\sum_i(S_i^z)^2+D^{\prime }\sum_i(S_i^z)^4
\end{equation}
where
\begin{eqnarray}
D &=&3B_2^0-[30S(S+1)-25]B_4^0  \label{rel} \\
D^{\prime } &=&35B_4^0  \nonumber
\end{eqnarray}
For $D,D^{\prime }>0$ spins of magnetic ions lie in the easy plane $xy$,
while for $D,D^{\prime }<0$ we have the easy axis $z.$ For $D>0,\;D^{\prime
}<0$ a first-order transition takes place between the easy plane (which is
favored by second-order anisotropy) and easy axis (which is favoured by
large $|D^{\prime }|$). We consider only the case $D<0,\;D^{\prime }>0$
where the cone phase occurs at intermediate values of $D/(2D^{\prime }S^2)$,
so that spin orientation direction makes the angle $\theta $ with the $z$%
-axis and the orientational phase transitions are of the second order. This
is the case for Gd and also for Ho, Er in low-temperature phases.

In the phenomenological approach it is supposed (see, e.g., Refs. \cite
{Levitin,Coqblin})
\begin{equation}
F=F_{\text{is}}+D(T)(S\cos \theta )^2+D^{\prime }(T)(S\cos \theta )^4
\end{equation}
where $F_{\text{is}}$ is the isotropic ($\theta $-independent) part of the
free energy. Then we obtain by minimization of $F$%
\begin{equation}
\cos ^2\theta (T)=-\frac{D(T)}{2D^{\prime }(T)S^2}  \label{ct}
\end{equation}
so that at the point where $D(T)=0$ the spins become directed in the $xy$
plane while at $|D(T)|\geq 2D^{\prime }(T)S^2$ they are aligned along the $z$%
-axis. The temperature dependence of $D(T)$ is supposed to have the form
\begin{equation}
D(T)=2D^{\prime }S^2(T_1-T)/(T_2-T_1)  \label{PhT}
\end{equation}
with $D^{\prime }(T)>0.$ Thus at $T=T_1$ the transition from the easy-plane
to cone structure takes place, while at $T=T_2$ the transition from the cone
to easy-axis structure occurs. At the same time, Zener's \cite{Zener} resut
for the temperature dependence of anisotropy constants in an axially
symmetric state with $\theta =0$ has the form
\begin{equation}
B_l^0(T)=B_l^0M^{l(l+1)/2}  \label{Zener}
\end{equation}
where $M=\langle \widetilde{S}^z\rangle /S$ is the relative magnetization,
and $D(T),$ $D^{\prime }(T)$ are determined by the same relations (\ref{rel}%
) with $B_l^0\rightarrow B_l^0(T)$. As pointed in Refs. \cite
{Tb,TbDy1,GdTbDy,TbDy2}, the temperature dependences of anisotropy constants
have a more complicated form for the cone structures with $\theta >0$ (in
fact, only the case $\theta =\pi /2$ is discussed in Refs. \cite
{Tb,TbDy1,GdTbDy,TbDy2}).

A systematic way of calculating temperature dependences of anisotropy
constants is the $1/S$-expansion which is considered in the next section.

\section{The $1/S$ expansion of the partition function}

The $1/S$-expansion developed here is slightly different from the standard
scheme of $1/S$-expansion \cite{Tb,GdTbDy,TbDy2} since it gives a
possibility to take into account exactly the kinematical relations between
powers of spin operators on each site. We use the coherent state approach
(see, e.g., Ref. \cite{Arovas}) to write down the partition function in the
form
\begin{equation}
{\cal Z}=\int D\mbox {\boldmath $\pi $}\exp \left\{ iS\int\limits_0^\beta
d\tau (1-\cos \vartheta )\frac{\partial \varphi }{\partial \tau }-\langle
\pi |{\cal H}|\pi \rangle \right\}   \label{Zp}
\end{equation}
where $\mbox {\boldmath $\pi $}$ is the unit-length vector, $\vartheta $ and
$\varphi $ are its polar and azimuthal angles respectively, $|\pi \rangle
=\exp (i\vartheta S^y+i\varphi S^z)|S\rangle $ are the coherent states ($%
S^z|S\rangle =S|S\rangle $). To construct the $1/S$-expansion we rotate the
coordinate system around the $y$-axis through the angle $\theta .$ The
Hamiltonian (\ref{H}) takes the form
\begin{eqnarray}
{\cal H} &=&-\frac J2\sum_{\langle ij\rangle }{\bf S}_i{\bf S}_j \\
&&\ \ \ +\sum_i\sum_{l,m}\sum_{m^{\prime }=-l}^lB_l^m\sqrt{\frac{%
(l+m)!(l-|m^{\prime }|)!}{(l-m)!(l+|m^{\prime }|)!}}\frac{A_l^m}{%
A_l^{|m^{\prime }|}}d_{mm^{\prime }}^l(\theta )(\widetilde{O}_l^{|m^{\prime
}|})_i  \nonumber
\end{eqnarray}
where $d_{mm^{\prime }}^l(\theta )$ are the Wigner matrices of the rotation
group irreducible representation,
\begin{equation}
A_l^m=\frac{(l-m)!}{(l+[m]-1)!!}\frac 1{K_l^m}
\end{equation}
($[m]=m$ for $m$ even and $[m]=m+1$ for $m$ odd), for $l\leq 4$ we have $%
K_l^m=1,$ and the tilde sign here and hereafter is referred to the rotated
coordinate system. Since the partition function (\ref{Zp}) is invariant
under rotation of the states $|\pi \rangle ,$ it is convenient to use the
coherent states defined in the same coordinate system, i.e., $|\widetilde{%
\pi }\rangle =\exp (i\vartheta \widetilde{S}^y+i\varphi \widetilde{S}^z)|%
\widetilde{S}\rangle \;$with $\widetilde{S}^z|\widetilde{S}\rangle =S|%
\widetilde{S}\rangle .$ The advantage of using the coherent states is the
simple form of the averages of the tensor operators (\ref{Tens}) over $|%
\widetilde{\pi }\rangle .$ By direct calculation we obtain
\begin{equation}
\langle \widetilde{\pi }|\widetilde{O}_l^m|\widetilde{\pi }\rangle
=S_lA_l^mP_l^m(\cos \vartheta )\cos m\varphi   \label{Av}
\end{equation}
where $P_l^m(x)$ are the associated Legendre polynomials, the factors $%
S_l=S(S-1/2)...[S-(l-1)/2]$ take into account properly the kinematical
relations on each site. In particular, the second-order anisotropy term
vanishes for $S=1/2,$ and the fourth-order for $S=1/2,1,3/2,$ as it should
be (unlike the results of boson representations in Refs.\cite
{Tb,GdTbDy,TbDy2}). Using (\ref{Av}) we obtain for the case $%
B_l^m=B_l^0\delta _{m0}$ under consideration the result
\begin{eqnarray}
\langle \widetilde{\pi }|{\cal H}|\widetilde{\pi }\rangle  &=&-\frac{JS^2}%
2\sum_{\langle ij\rangle }\widetilde{\mbox {\boldmath $\pi $}}_i\widetilde{%
\mbox {\boldmath $\pi $}}_j  \label{Cs} \\
&&\ \ \ +\sum_i\sum_{l=2,4}\sum_{m=-l}^lS_lB_l^0A_l^0\frac{(l-|m|)!}{(l+|m|)!%
}P_l^{|m|}(\cos \theta )P_l^{|m|}(\cos \vartheta )\cos m\varphi   \nonumber
\end{eqnarray}
Further calculations are performed in the same line as in Ref. \cite{OurITF}%
. Representing $\cos \vartheta =\sqrt{1-\sin ^2\vartheta }$ and expanding in
$\sin \vartheta $ we obtain the $1/S$-expansion of the partition function.
It should be stressed that we retain the factors $S_l,$ as well as $S$%
-dependences in (\ref{rel}), non-expanded. By performing decouplings, terms
of third order are reduced to linear ones, and terms of fourth order to
quadratic ones. The requirement of absence of $\sin \vartheta $-linear terms
leads to the result for the cone angle $\theta $
\begin{equation}
\cos ^2\theta =\frac 37\left[ 1-X+Y-\frac 1{10}\frac{B_2^0S_2}{B_4^0S_4}%
\left( 1-\frac 7{2S}+6X+Y\right) \right]   \label{Teta}
\end{equation}
where
\begin{eqnarray}
X &=&\langle \pi _{xi}^2\rangle \equiv \langle \sin ^2\vartheta \cos
^2\varphi \rangle =\sum_{{\bf q}}\frac{J_0-J_{{\bf q}}}{2E_{{\bf q}}}\coth
\frac{E_{{\bf q}}}{2T},  \nonumber \\
\;Y &=&\langle \pi _{yi}^2\rangle \equiv \langle \sin ^2\vartheta \sin
^2\varphi \rangle =\sum_{{\bf q}}\frac{J_0-J_{{\bf q}}+\Delta _0/S}{2E_{{\bf %
q}}}\coth \frac{E_{{\bf q}}}{2T},
\end{eqnarray}
and the ``bare'' magnon spectrum reads
\begin{eqnarray}
E_{{\bf q}} &=&S\sqrt{(J_0-J_{{\bf q}})(J_0-J_{{\bf q}}+\Delta _0/S)}
\label{Eq} \\
\Delta _0 &=&2\left[ 3B_2^0S_2\cos 2\theta -10B_4^0S_4(28\cos ^4\theta
-27\cos ^2\theta +3)\right] ,  \nonumber
\end{eqnarray}
$\Delta _0$ being the energy gap. The corrections in (\ref{Teta}) can be
collected into powers in the same way as in Refs.\cite{Rast2,OurSSWT} to
obtain the correct description of thermodynamics at not too low
temperatures. (In the presence of higher-order anisotropy this is essential
since the coefficients at $X,Y$ increase as $\sim l^2/2$ with anisotropy
order.) Then we have
\begin{equation}
\cos ^2\theta =\frac 37\frac{Z_X}{Z_Y}\left[ 1-\frac 1{10}\frac{B_2^0(T)S_2}{%
B_4^0(T)S_4}\right]   \label{Teta1}
\end{equation}
where
\begin{equation}
B_2^0(T)=Z_X^2Z_YB_2^0,\;B_4^0(T)=Z_X^9Z_YB_4^0  \label{AnisRen}
\end{equation}
are the temperature-renormalized anisotropy constants,
\begin{equation}
Z_X=1+\frac 1{2S}-X,\;Z_Y=1+\frac 1{2S}-Y
\end{equation}
The relations (\ref{AnisRen}) extend the results of Refs. (\cite
{Tb,TbDy1,GdTbDy,TbDy2}) to the case where spins make a non-zero angle with
the $z$-axis. The renormalized gap has the form
\begin{eqnarray}
\Delta  &=&6\cos 2\theta \,B_2^0S_2-20B_4^0S_4\left[ 3(1-7\widetilde{X}%
)-3\,\cos ^2\theta \,\left( 9-56\widetilde{X}-7\widetilde{Y}\right) \right.
\nonumber \\
&&\ \ \ \ \ \ \left. +28\,\cos ^4\theta \,\left( 1-6\widetilde{X}-\widetilde{%
Y}\right) \right] -196\sin ^2\theta \cos ^2\theta   \nonumber \\
&&\ \ \ \ \ \ \ \times \sum_{k,\omega _n}\left[ \frac{3B_2^0S_2(J_0-J_{{\bf k%
}})-10B_4^0(S_4/S)\Delta _0\cos ^2\theta }{\omega _n^2+S^2(J_0-J_{{\bf k}%
})(J_0-J_{{\bf k}}+\Delta _0/S)}\right] ^2  \label{SpRen}
\end{eqnarray}
where $\widetilde{X}=X-1/(2S),$ $\widetilde{Y}=Y-1/(2S)$. After introducing
the temperature-renormalized second- and fourth order anisotropy parameters $%
D(T)$ and $D^{\prime }(T),$
\begin{eqnarray}
D(T)S^2 &=&3B_2^0(T)S_2-30B_4^0(T)S_4,  \label{rel1} \\
D^{\prime }(T)S^4 &=&35B_4^0(T)S_4\left( Z_Y/Z_X\right) ,  \nonumber
\end{eqnarray}
the expression for $\cos \theta $ coincides with that of the
phenomenological theory (\ref{ct}). Collecting again corrections in (\ref
{SpRen}) into powers, we obtain for the renormalized gap in the notations (%
\ref{rel1}) the expression
\begin{eqnarray}
\Delta  &=&2D(T)S^2\cos 2\theta +2D^{\prime }(T)S^4\cos ^4\theta +6D^{\prime
}(T)S^4\sin ^2\theta \cos ^2\theta   \nonumber \\
&&\ \ \ \ -196\sin ^2\theta \cos ^2\theta \sum_{k,\omega _n}\left[ \frac{%
3B_2^0S_2(J_0-J_{{\bf k}})-10B_4^0(S_4/S)\Delta _0\cos ^2\theta }{\omega
_n^2+S^2(J_0-J_{{\bf k}})(J_0-J_{{\bf k}}+\Delta _0/S)}\right] ^2
\end{eqnarray}
which also coincides with that obtained in the phenomenological theory
except for the last term. Note that at $\theta >0$ the renormalizations (\ref
{rel1}) are present even at $T=0,$ which should be taken into account when
treating experimental data.

\section{Orientational phase transitions}

Now we can pass to description of possible orientational phase transitions.
Consider first the case of a small enough constant $B_4^0$ (or,
equivalently, $D^{\prime }$), so that $\cos ^2\theta (0)\;$ is close to
unity. Then $\cos ^2\theta (T)$ increases with temperature and there occurs
a transition to the easy-axis phase at the point determined by
\begin{equation}
\frac 37\frac{Z_X}{Z_Y}\left[ 1-\frac 1{10}\frac{B_2^0(T)S_2}{B_4^0(T)S_4}%
\right] =1  \label{PhT1}
\end{equation}
In the opposite case of a large enough $B_4^0,$ $\cos ^2\theta (0)$ is
small, and $\cos ^2\theta (T)$ decreases with temperature, so that at the
point where
\begin{equation}
B_2^0(T)S_2=10B_4^0(T)S_4  \label{PhT2}
\end{equation}
a phase transition to the easy-plane phase occurs. Thus one can expect that
there exists the critical value $\theta _c$: for $\theta _0=\theta
(0)<\theta _c$ we have a decrease of $\theta (T)$ with $T$ and the phase
transition from cone to easy-axis phase, while for $\pi /2>\theta _0>\theta
_c$ we have an increase of $\theta (T)$ with $T$ and the phase transition
from cone to easy-plane phase. The numerical computations for the simple
cubic lattice (see Fig.1) yield $\theta _c\simeq 50^{\circ }$. Fig.2 shows
the corresponding temperature dependences of the anisotropy constants $D(T),$
$D^{\prime }(T)$. For simplicity, $J_{{\bf q}}$ is taken for the simple
cubic lattice.

The phase transitions described by Eqs. (\ref{PhT1}) and (\ref{PhT2}) are
analogous to those in the phenomenological theory of Ref. \cite{Levitin}
that occur at $D(T)=0$ and $D(T)=-2D^{\prime }(T)S^2,$ respectively.
However, unlike the phenomenological approach, microscopical consideration
leads to either cone to easy-axis or cone to easy-plane transition with
increasing temperature, depending on the zero-temperature value of $\theta .$
At the same time, the transition from the easy-plane to easy-axis structure
(through the intermediate cone phase) cannot be explained by purely magnetic
renormalizations of anisotropy constants.

The result (\ref{Teta}) gives the mean-field values of the critical
exponents (e.g., $\beta =1/2$) for both the ground-state and temperature
orientational phase transitions. Unlike the systems discussed in Ref. \cite
{OurITF}, the system under consideration has the dynamical critical exponent
$z=2$ (i.e., single excitation mode with nearly quadratic dispersion is
present). Thus the upper critical dimensionality for the ground-state QPT is
$d_c^{+}=4-z=2.$ In this respect, the system is analogous to $XY$ model in
the transverse magnetic field \cite{XY-TF}. A characteristic feature of such
systems is the mean-field behavior of critical exponents both above and
below the critical dimensionality. For (hypothetical) systems with $d=2$
logarithmic corrections to ground-state properties near QPT are present
(see, e.g., Ref. \cite{Sachdev}). At the same time, the upper critical
dimensionality for the temperature phase transition is $d_{cT}^{+}=4,$ and
at $d<d_{cT}^{+}$ the temperature-transition critical exponents differ from
their mean-field values.

Now we discuss the experimental situation. In Gd (see, e.g., Refs. \cite
{Coqblin,JensenBook}) the orientational phase transition from cone-phase to
easy-axis phase is observed at $T_c=240$K. The temperature dependence of the
cone angle at $T<T_c$ (and also of magnetic anisotropy constants) is
non-monotonous, unlike the results obtained in Sect 2. This complicated
situation is connected with the absence of orbital momentum and smallness of
anisotropy in gadolinium.

In holmium the low-temperature phase is conical spiral one, the angle of the
cone changing from $\approx 80^{\circ }$ to $90^{\circ }$ in the temperature
interval $0-20K.$ The spiral angle makes up about $30^{\circ }.$ Since the
sixth-order anisotropy is important, we use the Hamiltonian \cite{Jensen}
\begin{equation}
{\cal H}_{\text{Ho}}={\cal H}+B_6^0\sum_i(O_6^0)_i+B_6^6\sum_i(O_6^6)_i
\label{HHo}
\end{equation}
The hcp lattice is not of a Bravais type. However, if we neglect the optical
mode (which is possible at $T\ll T_N=133$ K) one can put (see, e.g., Ref.
\cite{Coqblin})
\begin{eqnarray}
J_{{\bf q}} &=&2J\left[ \cos q_x+2\cos (q_x/2)\cos (\sqrt{3}q_y/2)\right]
\nonumber \\
&&\ \ \ \ +2J^{\prime }\cos \frac{q_z}2\left| \exp (iq_y/\sqrt{3})+2\cos
\frac{q_x}2\exp (-iq_y/2\sqrt{3})\right|
\end{eqnarray}
The parameters of the Hamiltonian were taken from Ref. \cite{Jensen}: $%
J=0.65 $K$,$ $J^{\prime }=0.6J,\;B_2^0=0.35$K,\ $B_4^0=0,\;B_6^0=-1.1\cdot
10^{-5}$K, $B_6^6=1.07\cdot 10^{-4}$K (note that our value of $B_2^0$
includes also renormalization due to dipolar anisotropy). For simplicity, we
restrict ourselves to a collinear magnetic structure (this is justified by
that the spiral angle in the rare earths is small, especially at low
temperatures). The calculations with the Hamiltonian (\ref{HHo})\ are
completely analogous to those in the previous Section. Calculated dependence
of the cone angle is compared with the result of the mean-field
approximation and experimental data in Fig.3. One can see that our results
improve somewhat those of the mean-field theory where the temperature
dependence of the anisotropy constants is given by (\ref{Zener}).

To conclude, we have formulated a consistent spin-wave approach to
description of thermodynamic properties of anisotropic magnets at low
temperatures. The renormalizations of the anisotropy constants and spin-wave
spectrum for an arbitrary cone angle are calculated. This gives a
possibility to describe the orientational phase transition between the cone
and plane phases.

We are grateful to J.Jensen for comments concerning the experimental
situation in holmium.

{\sc Figure captions}

Fig.1. The theoretical temperature dependences of the cone angle $\theta (T)$
for $S=7/2$ and different values of second-order anisotropy: $D/J=0.004;$ $%
0.005;$ $0.006$ from upper to lower curve. The value of $D^{\prime }/J$ is $%
3.7\cdot 10^{-4}.$

Fig.2. The temperature dependences of the anisotropy constants $D(T),$ $%
D^{\prime }(T)$ corresponding to Fig.1.

Fig.3. Calculated dependences of the cone angle in the mean-field
approximation (short-dashed line) and renormalized spin-wave theory (RSWT,
long-dashed line) as compared with experimental points for holmium (Refs.
\cite{Coqblin,HoEx1}).


\begin{thebibliography}{*}
\bibitem[*]{}  Electronic address: Valentin.Irkhin@imp.uran.ru

\bibitem{Levitin}  Belov K P, Zvezdin A\ K, Kadomtzeva A\ M and Levitin P\ Z
1979 {\it Orientational transitions in the rare-earth magnets} (Moscow,
Nauka) [in Russian]

\bibitem{Coqblin}  Coqblin B 1977 {\it The Electronic Structure of
Rare-Earth Metals and Alloys: the Magnetic Heavy Rare-Earths} (Academic
Press, London)

\bibitem{Tb}  Brooks M S S Goodings D A and Ralph H\ I 1968{\it \ J.Phys.C }%
{\bf 1} 132

\bibitem{TbDy1}  Brooks M\ S\ S 1970 {\it Phys.Rev.B} {\bf 1} 2257

\bibitem{GdTbDy}  Lindgard P A and Danielsen O 1975, {\it Phys.Rev.B }{\bf 11%
} 351

\bibitem{TbDy2}  Jensen J 1975 {\it J.Phys.C} {\bf 8} 2769

\bibitem{OurITF}  Irkhin V Yu and Katanin A A 1998 {\it Phys.Rev. B }{\bf 57}
379

\bibitem{Zener}  Zener C 1954 {\it Phys.Rev. }{\bf 96} 1335

\bibitem{Arovas}  Auerbach A 1994 {\it Interacting Electrons and Quantum
Magnetism} (Springer-Verlag, New York)

\bibitem{Rast2}  Rastelli E, Tassi A and Reatto L 1974, {\it J.Phys.C }{\bf 7%
}, 1735.

\bibitem{OurSSWT}  Irkhin V Yu, Katanin A A and Katsnelson M I, {\it Phys.
Rev.B}, in press.

\bibitem{XY-TF}  Gerber P R and Beck H 1977, {\it J.Phys.C }{\bf 10}, 4013 %
\nonum Kopec T K and Kozlowski G 1983, {\it Phys.Lett. A }{\bf 95}, 104

\bibitem{Sachdev}  Sachdev S, Senthil T and Shankar R 1994, {\it Phys. Rev. B%
} {\bf 50}, 258.

\bibitem{JensenBook}  Jensen J and Mackintosh A R 1991 {\it Rare Earth
Magnetism: Structures and Excitations} (Oxford University Press, Oxford)

\bibitem{Jensen}  Larsen C C, Jensen J and Mackintosh A R 1987, {\it %
Phys.Rev.Lett.} {\bf 59}, 712

\bibitem{HoEx1}  Koehler W C, Cable J W, Wilkinson M K and Wollan E O 1966
{\it Phys.Rev.} {\bf 151}, 414
\end{thebibliography}
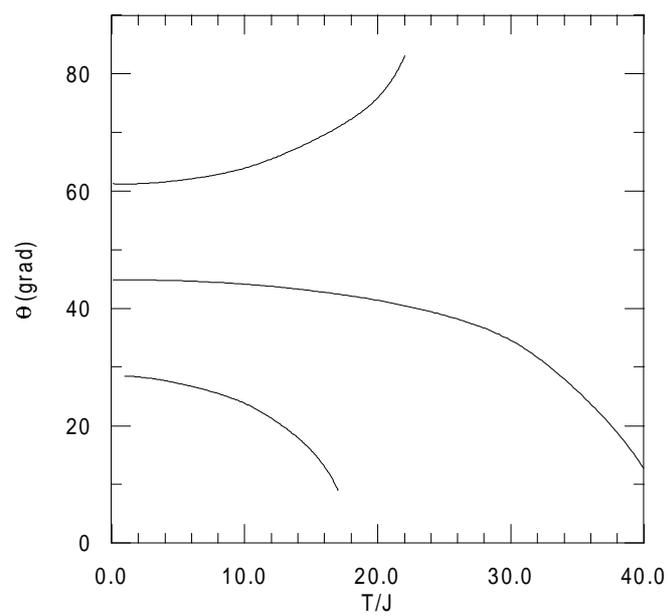
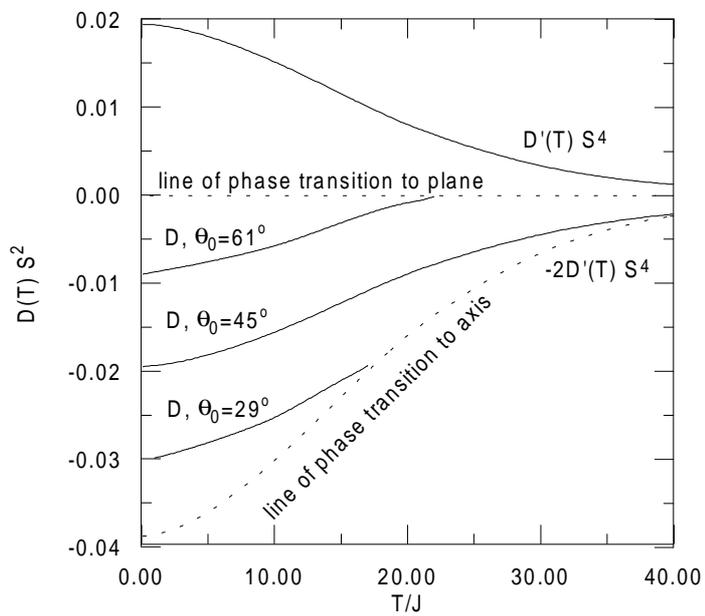
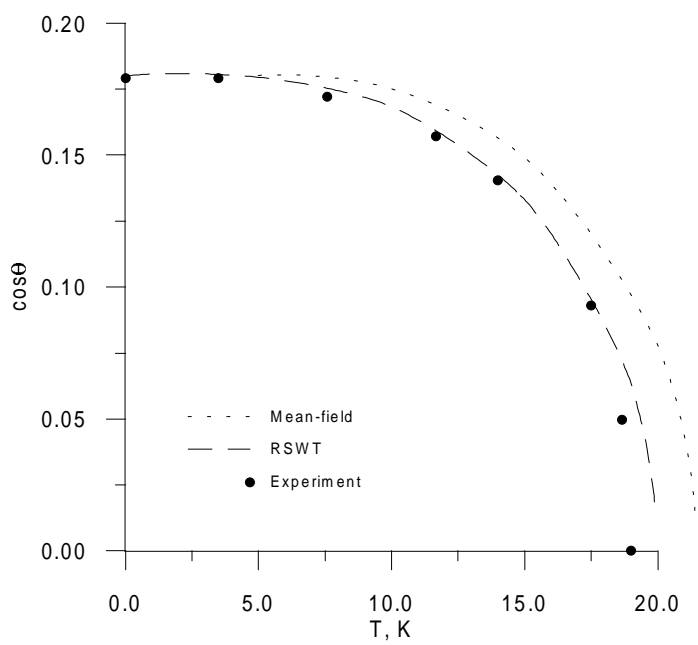
\end{document}